# Residual neural-field ptychography for dose-efficient electron, X-ray, and optical nanoscopy

Qianhao Zhao[1,†], Zhixuan Hong[1,†], Ruihai Wang[1], Tianbo Wang[1], Lingzhi Jiang[1], Qiong Ma[1], Peng-Han Lu[2], Rafal E. Dunin-Borkowski[2], Andrew Maiden[3,4], and Guoan Zheng[1,*]

[1]Department of Biomedical Engineering, University of Connecticut, Storrs, CT 06269, USA
[2]Ernst-Ruska Centre for Microscopy and Spectroscopy with Electrons, Forschungszentrum Jülich GmbH, Jülich 52425, Germany
[3]Department of Electronic and Electrical Engineering, University of Sheffield, Sheffield, SYK S1 3JD, UK
[4]Diamond Light Source, Harwell, Oxfordshire OX11 0DE, UK
[†]These authors contributed equally to this work
[*]Email: guoan.zheng@uconn.edu

**Abstract:** Ptychography spans from sub-angstrom to meter scales yet suffers from convergence instability and excessive data redundancy. Here we introduce self-correcting residual neural fields as a dose-efficient framework for electron, X-ray, and optical ptychography. Unlike approaches that split complex fields, our complex-valued architecture employs holomorphic phasor activation $e^{i\omega z}$ to preserve intrinsic phase-amplitude coupling. We reformulate reconstruction as residual learning, where the network learns only corrections to physical priors rather than complete wavefields. By embedding the physical model as a differentiable layer within the network, we enable end-to-end automatic differentiation where experimental parameters are jointly corrected alongside the neural fields. We validate our scheme across conventional, near-field, coded, and Fourier ptychography and achieve record-breaking lensless resolution of 244-nm linewidth with visible light. Extending to electron wavelengths, we reveal synaptic connectivity in brain sections with superior performance over conventional approaches. Our framework provides a solution for high-throughput, dose-efficient nanoscopy across the electromagnetic spectrum.

## Introduction

Ptychography has established itself as a transformative computational imaging modality that operates across nearly nine orders of magnitude in length scale, from picometre-scale electron microscopy to metre-scale optical sensing[1]. By replacing physical lens elements with computational reconstruction, the technique has enabled deep-sub-angstrom resolution in electron microscopy[2,3,4] and non-destructive three-dimensional tomography of integrated circuits using hard X-rays[5,6]. In addition, new modalities in optical domain such as Fourier ptychography[7] and coded ptychography[8] have successfully overcome the fundamental trade-off between resolution and field of view[9,10], enabling the synthesis of gigapixel-scale images with high space-bandwidth products for biomedical and industrial applications[11,12,13,14].

Despite these advances, conventional phase retrieval algorithms impose acquisition constraints that limit ptychography's throughput and applicability to radiation-sensitive specimens. Standard iterative solvers, such as various ptychographical iterative engines[15,16], weighted average of sequential projections[17], quasi-Newton methods[18], or least-squares solvers[19], rely heavily on data redundancy to break the mathematical ambiguity between the object and the probe[1,20]. This requirement typically necessitates overlap ratios of 50% or more between adjacent scan positions, which imposes limitations on temporal resolution and exposes radiation-sensitive specimens to excessive dose[2,21,22,23]. In electron ptychography, knock-on damage from elastic scattering directly displaces atoms and inelastic interactions cause radiolysis and specimen heating; similarly, hard X-rays induce structural changes through absorbed dose. Beyond data redundancy, these algorithms often rely on knowledge of experimental parameters, including scan positions, diffraction distances, illumination wavelength, detector geometry, and aberration coefficients, which are often imprecisely known or drift during acquisition. While conventional algorithms can refine select parameters through dedicated update rules, jointly optimizing experimental uncertainties remains challenging, often requiring specialized algorithmic extensions for each parameter type and careful tuning to avoid instability. Furthermore, these pixel-grid-based solvers frequently confront convergence stagnation when faced with noise, initialization problems, raster-scan pathology, and phase wrapping issue of thick objects.

Recently, implicit neural representations, or neural fields, have emerged as a promising alternative by parameterizing the object as a continuous function within a multilayer perceptron[14, 24, 25, 26, 27, 28, 29, 30, 31, 32, 33, 34, 35]. However, conventional coordinate-based networks are often ill-suited for the physics of coherent diffraction. Most existing architectures treat the complex wavefield as two independent real-valued channels, either amplitude-phase or real-imaginary separation, driven by standard activation functions like rectified linear unit (ReLU). This 'split-channel' approach treats amplitude and phase as independent quantities, requiring redundant parameterization to capture what should be a unified complex field and preventing the network from exploiting the physical correlations between amplitude and phase. The amplitude-phase representation further introduces non-differentiable discontinuities at $2\pi$ phase-wrapping boundaries, producing artifacts and trapping optimization in local minima. Moreover, current neural fields learn the entire wavefield *ab initio* from random initialization, discarding prior information routinely available in ptychographic experiments, such as probe beams estimated from defocus and convergence angle in conventional ptychography[36], pre-calibrated coded surfaces in coded ptychography[10, 37], and pupil functions in Fourier ptychography[38, 39]. Without incorporating these priors, neural field optimization frequently fails to converge, as the ill-posed inverse problem provides insufficient constraints to guide the network away from local minima, especially when the object and probe contain many phase wraps[40].

Here, we introduce complex residual neural fields with end-to-end differentiable physics, a unified physics-informed framework that addresses these limitations across electron, X-ray, and optical ptychography. Departing from standard split-channel architectures, we utilize a fully complex-valued neural network driven by a holomorphic phasor activation function. This activation function mirrors the fundamental oscillatory nature of diffracting wavefields, ensuring that the network naturally preserves the coupling between amplitude and phase. Crucially, we reformulate the inverse problem from *ab initio* reconstruction to a residual learning task, inspired by the deep residual learning (ResNet) framework in computer vision[41]. The core insight of ResNet is that it is easier to optimize residual mapping than to learn an unreferenced function from scratch. We adapt this principle to wave physics: by leveraging pre-calibrated probes or initial estimates as hard physical priors ($\psi_{\text{initial}}$), our network focuses exclusively on learning the complex residual $\Delta\psi = \psi_{\text{true}} - \psi_{\text{initial}}$. This residual formulation leverages prior information routinely available in ptychographic experiments, transforming the ill-posed inverse problem into a well-conditioned one where the optimizer focuses on high-frequency corrections rather than reconstructing global structures from scratch. Since these residuals are inherently sparse, we further enforce wavelet-domain sparsity to significantly reduce data redundancy requirements. Beyond the object representation, we extend differentiability to the measurement process itself by embedding the imaging model as a differentiable layer within the framework. This enables end-to-end automatic differentiation[42, 43, 44, 45] where experimental parameters are jointly optimized alongside the neural fields, achieving rigorous self-correction from diffraction data.

Our scheme is validated with conventional, near-field, coded, and Fourier ptychography from optical to electron wavelengths. With coded ptychography, we resolve 244-nm linewidths on the resolution target, achieving the highest lensless resolution with visible light[1, 40]. Extending to electron wavelengths, we apply near-field electron ptychography to brain tissue sections for the first time, visualizing synaptic connectivity with superior reconstruction quality compared to conventional phase retrieval approaches. Furthermore, our scheme eliminates raster-grid pathology[46], enabling artifact-free reconstruction with simple raster scans. By unifying wave physics with residual learning, the reported framework provides a robust foundation for high-throughput, data-efficient nanoscopy across the electromagnetic spectrum.

## Results
**Residual neural fields with phasor activations and self-correcting differentiable layers**
Figure 1 shows the reported framework using fully complex-valued neural networks with end-to-end differentiable layers, designed to rigorously respect wave physics while enabling joint optimization of experimental parameters. This framework can accommodate diverse ptychographic configurations[1] including conventional, near-field, coded, and Fourier ptychography, each employing distinct forward models as detailed in Supplementary Note 1 and Supplementary Fig. S1. To efficiently capture high-frequency spatial details without the computational cost of deep networks, we first project the spatial coordinates $r = (x, y)$ onto a high-dimensional feature manifold using multiresolution hash encoding[26] (Fig. 1, top and bottom panels). For each resolution level $l$, the discrete coordinate indices are mapped to a feature vector via a hash function. These real-valued feature vectors are projected into the complex domain and processed by a multilayer perceptron (MLP) driven by a holomorphic phasor activation function:

$$\sigma(z) = exp\,(i\omega z) \qquad (1)$$

where $\omega$ is a frequency parameter. Crucially, this function satisfies the Cauchy-Riemann condition $\partial \sigma / \partial \bar{z} = 0$ (where $\bar{z}$ denotes the complex conjugate of $z$), confirming its holomorphic nature. Unlike split-channel activations such as complex ReLU, where dependence on the complex conjugate ($\partial \sigma / \partial \bar{z} \neq 0$) breaks the analytic structure, the phasor activation ensures that gradient updates strictly preserve the intrinsic coupling between amplitude and phase inherent to complex wavefields.

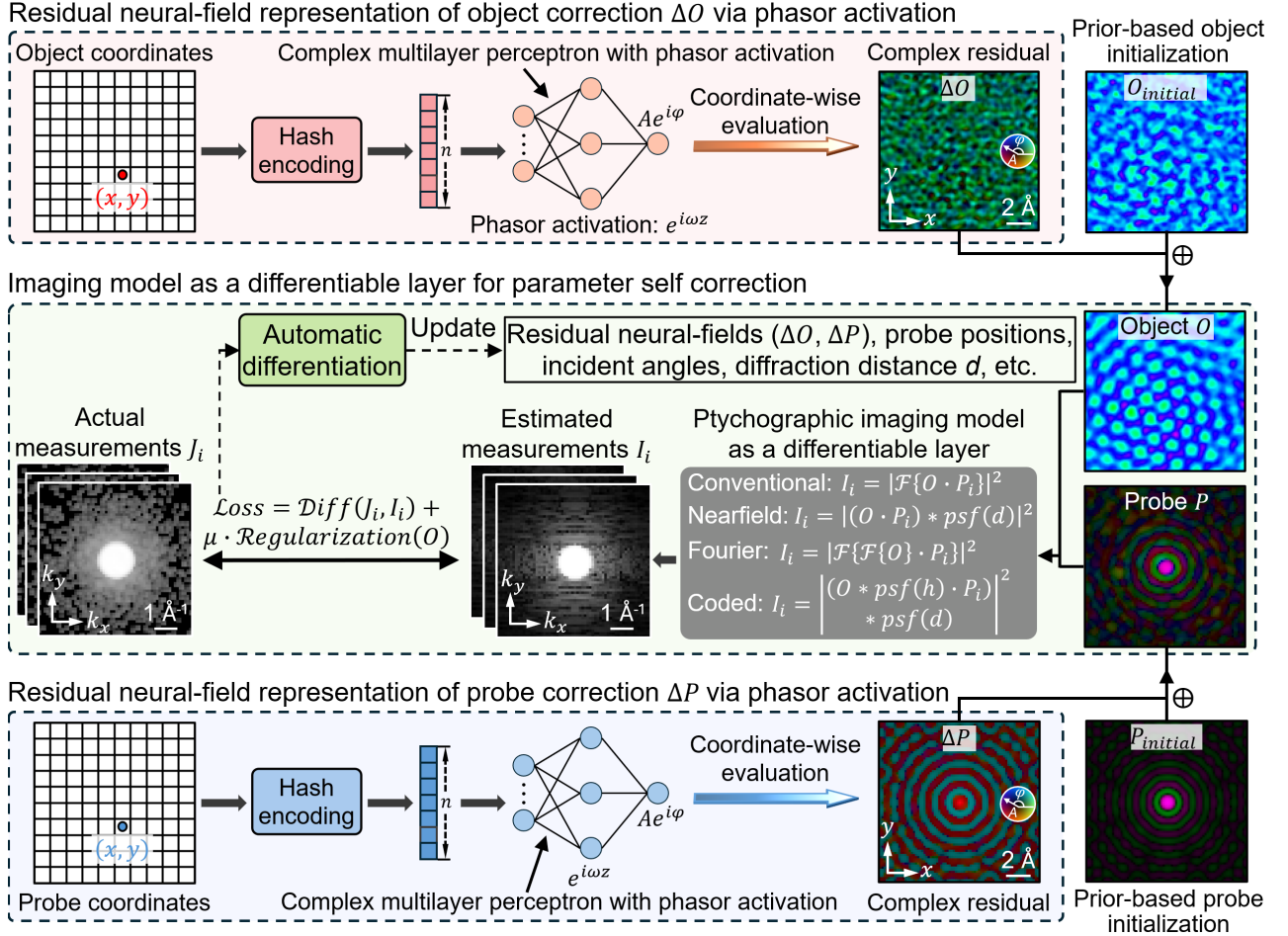

**Fig. 1 | Self-correcting residual neural fields with end-to-end differentiable layers.** The framework employs residual neural field representation for the complex object correction $\Delta O$ (top panel) and probe correction $\Delta P$ (bottom panel). In these neural field representations, spatial coordinates $(x, y)$ are mapped via hash encoding to a high-dimensional feature space, then processed by a complex multilayer perceptron with holomorphic phasor activation $e^{i\omega z}$ to produce the complex residuals $\Delta O$ and $\Delta P$. These complex residuals are summed with prior-based initializations $O_{initial}$ and $P_{initial}$ to yield the final object estimate $O$ and $P$. The imaging model is embedded as a differentiable layer within the framework (middle panel). This enables end-to-end automatic differentiation where gradients flow from the loss function through the imaging physics back to all learnable components: the residual neural fields ($\Delta O$ and $\Delta P$), scan positions $(x_i, y_i)$ in real space, incident wavevectors $(k_{xi}, k_{yi})$ in reciprocal space, diffraction distance $d$, etc. The loss function combines a data fidelity term with wavelet-domain sparsity regularization on the residuals, enabling robust reconstruction from reduced measurements. By treating system parameters as jointly learnable variables rather than fixed inputs, the framework achieves rigorous self-correction from diffraction measurements.

This activation extends the sinusoidal representation networks (SIREN) framework[28] to the complex domain. While SIREN employs $sin(\omega z)$ to capture high-frequency content in real-valued signals, our phasor activation operates on complex-valued inputs and weights, naturally encoding the oscillatory behaviour of wavefields. This formulation parameterizes the solution space as a smooth manifold of continuous, wave-like functions that intrinsically preserve amplitude-phase coupling. In Supplementary Note 2 and Supplementary Figs. S2-S3, we conducted a systematic comparison of complex-valued activation functions, including complex Tanh[47], complex ReLU[47], zReLU[48], modulus ReLU, and modReLU[49]. The reported phasor activation demonstrates superior performance compared to all these complex-valued activation functions.

To overcome the convergence stagnation typical of *ab initio* phase retrieval, we reformulate the inverse problem as a residual learning task that leverages prior information routinely available in ptychographic experiments. In ResNet[41], it has been shown that learning a residual mapping $\mathcal{F}(\mathbf{x}) = \mathcal{H}(\mathbf{x}) - \mathbf{x}$ is substantially easier than learning the unreferenced mapping $\mathcal{H}(\mathbf{x})$ directly. In our implementation, we anchor the reconstruction to physical priors $O_{\text{initial}}$ and $P_{\text{initial}}$ derived from calibrated probes, known optical parameters, or low-resolution estimates available in various ptychographic modalities. The complex neural fields $\mathcal{F}_{\theta_O}(r)$ and $\mathcal{F}_{\theta_P}(r)$ are tasked with learning the complex residual object $\Delta O$ and residual probe $\Delta P$ as follows:

$$\mathcal{F}_{\theta_O}(r) = \Delta O = O(r) - O_{\text{initial}}(r), \quad (2)$$
$$\mathcal{F}_{\theta_P}(r) = \Delta P = P(r) - P_{\text{initial}}(r), \quad (3)$$

where $O(\mathbf{r})$ and $P(\mathbf{r})$ are the final object and probe. This residual formulation ensures optimization begins near the physical prior, directing network capacity toward learning corrections rather than the global structures from scratch.

Another defining feature of our framework is to embed the physical imaging model as a differentiable layer within the computational graph (Fig. 1, middle panel). The operations in the forward model, i.e., Fourier transforms, element-wise multiplications, and Fresnel propagation, are implemented differentiably, enabling gradients to flow from the loss function through the imaging physics to all learnable parameters. This end-to-end differentiable architecture enables automatic differentiation with respect to both the neural field parameters and experimental variables that are conventionally treated as fixed inputs. The set of learnable parameters $\theta$ can encompass:

$$\theta = \{\theta_O, \theta_P, (x_i, y_i) \text{ or } (k_{xi}, k_{yi}), d, \lambda, (c_{1,1}, c_{2,0}, c_{2,-2}, c_{3,1}, \dots), \dots\} \quad (4)$$

where $\theta_O$ and $\theta_P$ are the weights of neural field representations of $\mathcal{F}_{\theta_O}$ and $\mathcal{F}_{\theta_P}$, $(x_i, y_i)$ are the scan positions in real space, $(k_{xi}, k_{yi})$ are the scan positions in reciprocal space for Fourier ptychography, $d$ is the diffraction distance, and $\lambda$ is the incident wavelength. For lens-based systems, the pupil function can be parameterized using Zernike polynomial coefficients $(c_{1,1}, c_{2,0}, c_{2,0}, c_{2,-2}, \dots)$ that describe optical aberrations[39], enabling joint optimization of aberration correction during reconstruction. Additional parameters such as the intensity fluctuation factors[50], the frequency parameter $\omega$ of the holomorphic phasor activation function, and regularization coefficients can also be included. This joint optimization achieves self-correction of system uncertainties.

The network is trained by minimizing a composite loss function that combines a data fidelity term with wavelet-domain sparsity regularization (Fig. 1, middle panel). The data fidelity term quantifies the discrepancy between the estimated intensity $I_i$ from the forward model and the measured intensity $J_i$, with options[1, 51] including absolute difference (L1), smooth absolute difference (L1-smooth), squared difference (L2), or gradient-domain formulations. The gradient-domain formulation provides robustness against low-frequency background variations and noises[9, 32, 52, 53]. The wavelet sparsity constraint exploits the compressibility of object recovery, enabling robust reconstruction from reduced measurements while suppressing non-physical artifacts and preserving sharp edges. We employ AdamW optimizer[54] for all experiments, which demonstrates superior convergence across various ptychographic modalities as detailed in Supplementary Note 3 and Supplementary Fig. S4.

**Universality across wavelengths and imaging modalities**
We validated the universality of the residual neural field framework by applying it to diverse experimental configurations across electron, X-ray, and optical regimes, encompassing conventional[55], near-field[56], coded[8, 57], and Fourier ptychography[7, 58] (Fig. 2). A key advantage of our residual approach is the proper integration of initialization priors for both object and probe. For conventional electron and X-ray ptychography, the probe prior can be estimated from the defocus distance or the beam convergence angle[59]. For coded and near-field ptychography, the coded modulation profile can be pre-calibrated using a weakly scattering reference specimen such as a blood smear slide[10].

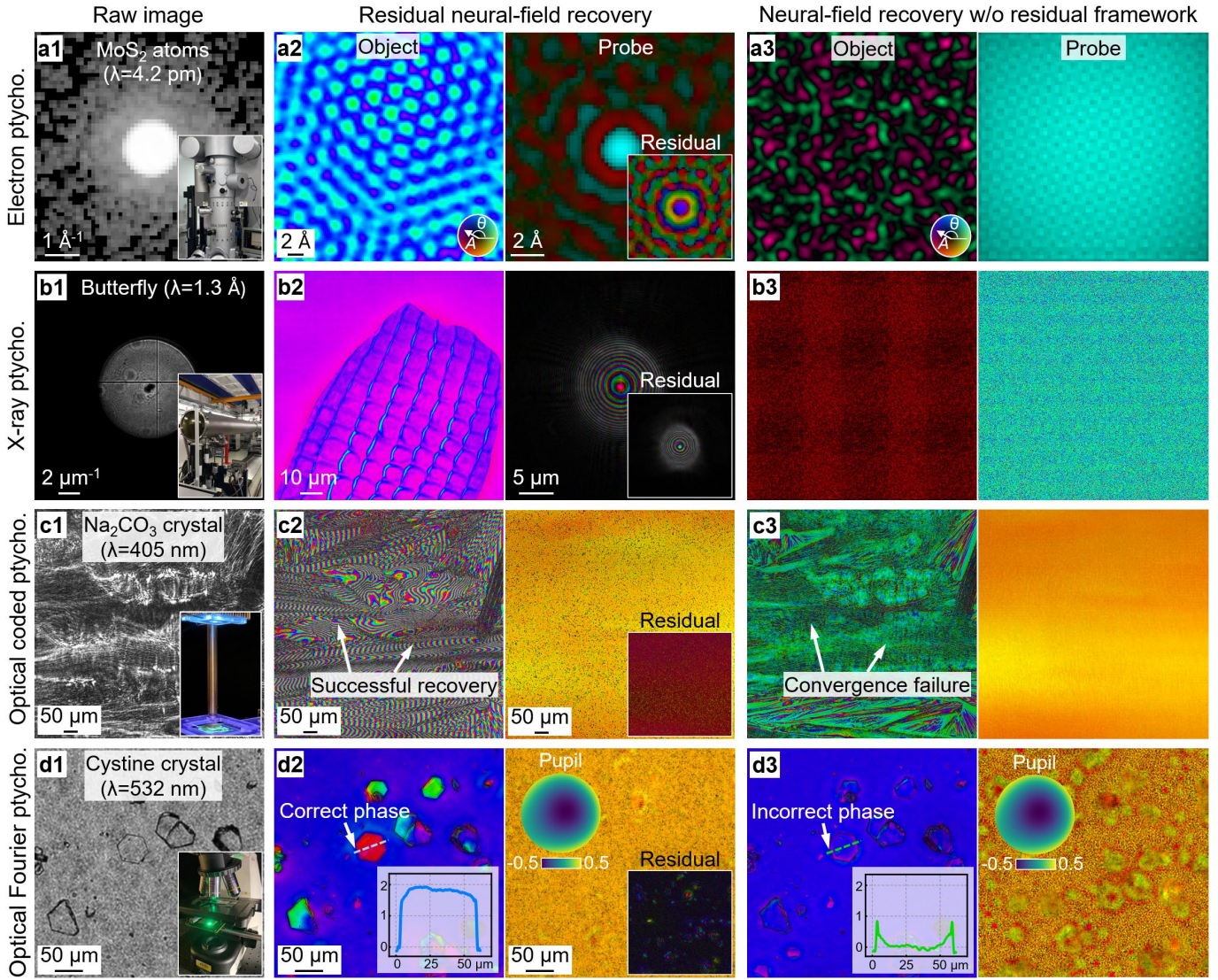

**Fig. 2 | Universality of the residual neural field framework across wavelengths and ptychographic modalities.** Each row presents a different imaging configuration with raw data (left column), residual neural-field recovery (middle column), and neural-field recovery without the residual framework (right column). The brightness of the recovered wavefields indicates amplitude ($A$), and hue indicates phase ($\theta$), as defined by the color wheel. Insets show experimental setups, recovered probe residuals, and line profile comparisons. **a**, Electron ptychography of MoS$_2$ atoms at $\lambda = 4.2$ pm, resolving the hexagonal atomic lattice[3]. The residual framework recovers both the atomic structure and the complex probe function, while the non-residual approach fails to resolve atomic features. **b**, Hard X-ray ptychography of a butterfly wing specimen at $\lambda = 1.3$ Å acquired at a synchrotron source[60]. The framework resolves sub-100-nm periodic microstructures and recovers the zone-plate probe, whereas the non-residual reconstruction produces featureless outputs. **c**, Lensless coded ptychography of Na$_2$CO$_3$ crystals acquired at $\lambda = 405$ nm. The residual framework successfully handles strong phase accumulation with multiple $2\pi$ wraps, while the non-residual approach exhibits convergence failure with phase-wrapping artifacts. **d**, Lens-based Fourier ptychography of cystine crystals acquired at $\lambda = 529$ nm. The residual framework yields correct quantitative phase values (flat line profile) and accurate pupil aberrations, whereas the non-residual approach produces incorrect phase with peaks at edges.

For electron ptychography (Fig. 2a), we applied our framework to diffraction data of MoS$_2$ atoms acquired at 80 eV ($\lambda = 4.2$ pm)[3]. The raw diffraction pattern (Fig. 2a1) shows the characteristic brightfield disk surrounded by high-angle scattering that encodes atomic-scale information. The residual neural-field recovery (Fig. 2a2) successfully resolved the atomic lattice with clear column visibility, while simultaneously recovering the complex probe residual exhibiting the phase structure characteristic of aberration-corrected electron optics. In contrast, neural-field recovery without the residual framework (Fig. 2a3) produced only artifacts with no discernible atomic features, demonstrating that the residual formulation with probe estimate is essential for convergence under the low signal-to-noise conditions inherent to electron imaging.

For hard X-ray ptychography (Fig. 2b), we applied our framework to diffraction data of a butterfly wing specimen acquired using synchrotron radiation at 9.7 keV ($\lambda$ =1.3 Å)[60]. The raw diffraction pattern (Fig. 2b1) exhibits a brightfield disk as a typical structure of coherent X-ray imaging. The residual framework (Fig. 2b2) recovered the intricate microstructure of the wing in high quality. The recovered probe displays concentric ring structures consistent with focusing X-ray optics, and the residual map confirms that the network learned the corrections to the prior. Without the residual framework (Fig. 2b3), the reconstruction failed completely, demonstrating the framework's inability to converge from random initialization when faced with a probe consisting of multiple phase wraps.

For lensless coded ptychography (Fig. 2c), we acquired wide-field-of-view diffraction measurements of $Na_2CO_3$ crystals at a wavelength of 405 nm. The coded surface transmission profile was pre-calibrated using a blood smear specimen[10]. The raw diffraction image (Fig. 2c1) displays complex interference patterns modulated by the coded surface. This sample presents a particularly challenging reconstruction problem: strong phase accumulation across crystal creates multiple $2\pi$ phase wraps that trap conventional blind reconstruction[1, 40, 58]. Applying the residual neural-field framework (Fig. 2c2), we successfully recover these phase wraps and the crystal morphology with quantitative phase values. Without the residual framework (Fig. 2c3), reconstruction exhibited severe convergence failure with prominent phase-wrapping artifacts throughout the field of view, confirming that the prior-constrained residual formulation is critical for breaking ambiguities in near-field diffraction geometries.

For lens-based Fourier ptychography (Fig. 2d), we acquired measurements of cystine crystals at a wavelength of 529 nm using angle-varied LED illumination combined with coded sensor detection. This spatially-coded configuration provides a uniform phase transfer function that enables true quantitative phase imaging[58], overcoming a fundamental limitation of regular Fourier ptychography which cannot recover slowly-varying phase information[40]. The coded surface prior was calibrated using a blood smear as a reference specimen[10]. The raw brightfield image in Fig. 2d1 exhibits low spatial resolution constrained by the NA of the objective lens. For both neural-field reconstructions with and without the residual framework, we parameterized the microscope pupil using 10 Zernike coefficients[39] within the differentiable imaging model, allowing these aberration coefficients to be jointly optimized during training. Applying our residual neural-field framework (Fig. 2d2), we recovered quantitative phase maps with correct values across the crystal structures, verified by line profile analysis showing the expected flat phase profile. In contrast, neural-field recovery without the residual framework (Fig. 2d3) produced incorrect phase values lacking the expected low-frequency variations, as these slowly varying phase components were erroneously absorbed into the coded surface profile during optimization.

The consistent failure of non-residual reconstruction across various modalities reveals the fundamental ill-posedness of joint object-probe retrieval without proper initialization. When reconstructing objects with substantial phase accumulation, such as crystals or thick biological specimens, the optimization faces inherent ambiguities: phase wraps in the object can be equivalently explained by spatial shifts of the probe, or slowly varying object phase can be absorbed into the probe or coded surface profiles. Without anchoring to physical priors, the network has no mechanism to resolve these degeneracies and converges to arbitrary solutions within the space of mathematically equivalent configurations. The residual formulation breaks these symmetries by constraining both object and probe to remain near their calibrated priors, ensuring that learned corrections represent genuine features rather than ambiguous redistributions of phase between reconstruction variables.

**Pushing resolution limits and data efficiency**

A key advantage of the residual formulation is its ability to direct neural network capacity specifically toward high-frequency details, enabling both enhanced resolution and reduced data acquisition requirements. Using a 405-nm laser diode and a lensless coded ptychography configuration, we demonstrated the resolution limits achievable with our framework in Fig. 3. The raw diffraction image (Fig. 3a) shows the characteristic speckle pattern produced by coded surface modulation of a USAF resolution target. With the residual framework, the object residual (Fig. 3b1) captures fine corrections to the initialization prior, and the final object (Fig. 3b2) displays sharp features down to the 244-nm linewidth corresponding to Group 11, Element 1. The coded surface residual (Fig. 3c1) and final coded surface (Fig. 3c2) show that the network maintains accurate separation between object and coded surface profile. Without the residual framework (Fig. 3d1-d2), the reconstruction resolved only 308-nm linewidths at Group 10, Element 5. The 244-nm linewidth demonstrated in Fig. 3b2 represents the highest resolution for lensless imaging with visible light.

To address the throughput bottleneck in ptychography, we incorporated wavelet-domain sparsity constraints into the loss function (Methods), exploiting the fact that the residual signal primarily consists of edge corrections and high-frequency textures that are naturally sparse in the wavelet basis. Supplementary Note 4 and Supplementary Fig. S5 detail the wavelet sparsity regularization and validate its effectiveness across different ptychographic modalities. We demonstrated this sparsity-constrained approach on severely undersampled datasets across electron, X-ray, and optical regimes in Fig. 4.

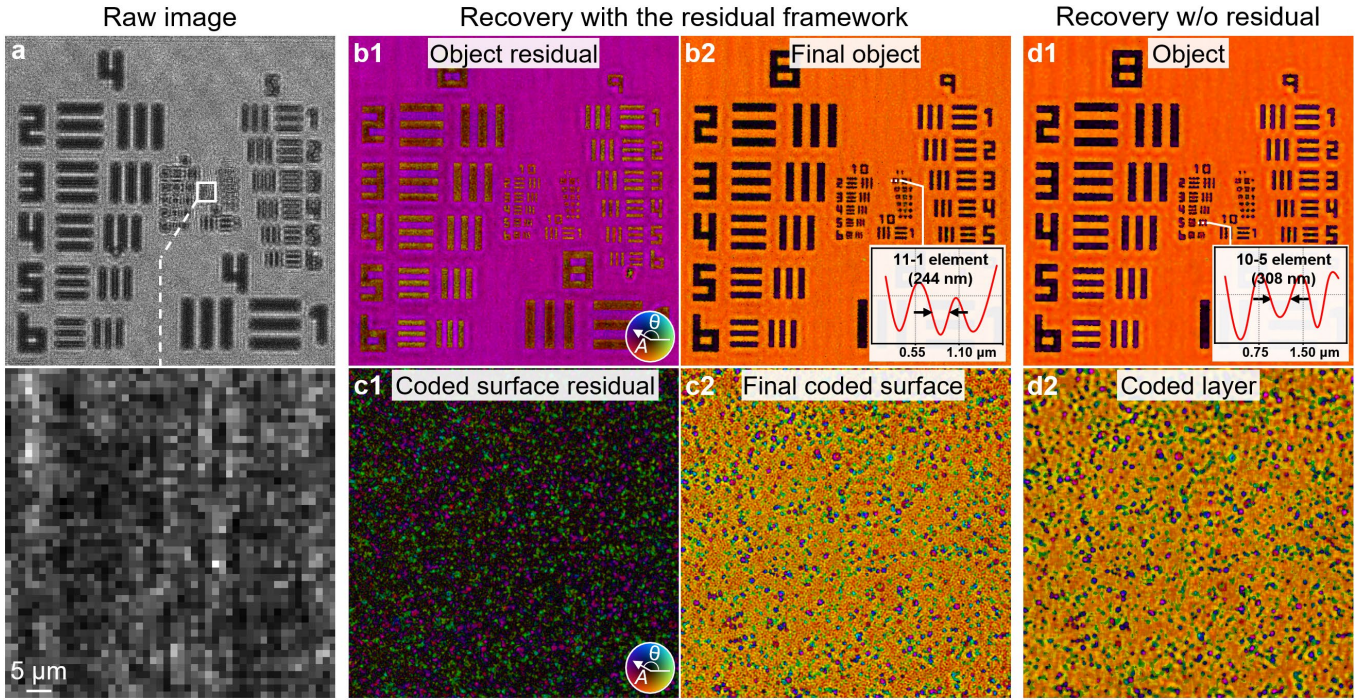

**Fig. 3 | Record resolution in lensless diffraction imaging with visible light. a**, Raw diffraction image of a USAF resolution target acquired via lensless coded ptychography at 405 nm, with zoomed view showing the Groups 8-11 region as in b2 and d1. **b1**, Object residual recovered by the complex MLP relative to the initialization prior. **b2**, Final reconstructed object obtained by summing the residual with the prior, resolving the 244-nm linewidth (Group 11, Element 1) as confirmed by line trace profile (inset). **c1**, Coded surface residual. **c2**, Final coded surface reconstruction showing recovery of the modulation pattern. **d1**, Object reconstruction without the residual framework, resolving only 308-nm linewidth (Group 10, Element 5) as shown in the intensity profile (inset). **d2**, Coded surface reconstruction without the residual framework. The achieved 244-nm linewidth represents the highest resolution demonstrated for lensless imaging with visible light. The brightness of the recovered wavefields indicates amplitude ($A$), and hue indicates phase ($\theta$), as defined by the color wheel in b1 and c1. Supplementary Fig. S2 provides a systematic comparison of six complex-valued activation functions on this resolution target, confirming that the holomorphic phasor activation achieves the highest resolution among all tested activations.

For near-field electron ptychography, we acquired diffraction data from mouse brain tissue sections prepared following established protocols to demonstrate reconstruction under sparse sampling conditions (Methods). With 50% fewer diffraction patterns, conventional iterative phase retrieval produced noisy reconstructions with poor contrast in both amplitude and phase channels. In contrast, the residual neural-field recovery with wavelet sparsity revealed synaptic structures and neural processes with substantially improved image quality, demonstrating the framework's utility for dose-sensitive biological specimens (Fig. 4a-b). This capability is important for high-resolution neural mapping, or connectomics, where the goal is to trace dense 3D wiring diagrams of brain circuits.

Similarly, for atomic-resolution electron ptychography of $MoS_2$ with 50% data reduction, conventional methods exhibited convergence failure, producing only noise with no discernible atomic features. Our framework successfully resolved the atomic lattice, confirming that the wavelet sparsity constraint acts as an effective physics-informed regularizer under data-limited conditions (Fig. 4c-d). For soft X-ray ptychography with 85% fewer acquisitions (Fig.

4e-f), conventional phase retrieval yielded low-contrast reconstructions with artifacts obscuring the nanogold sample features[61]. The residual framework maintained structural fidelity with clear feature boundaries.

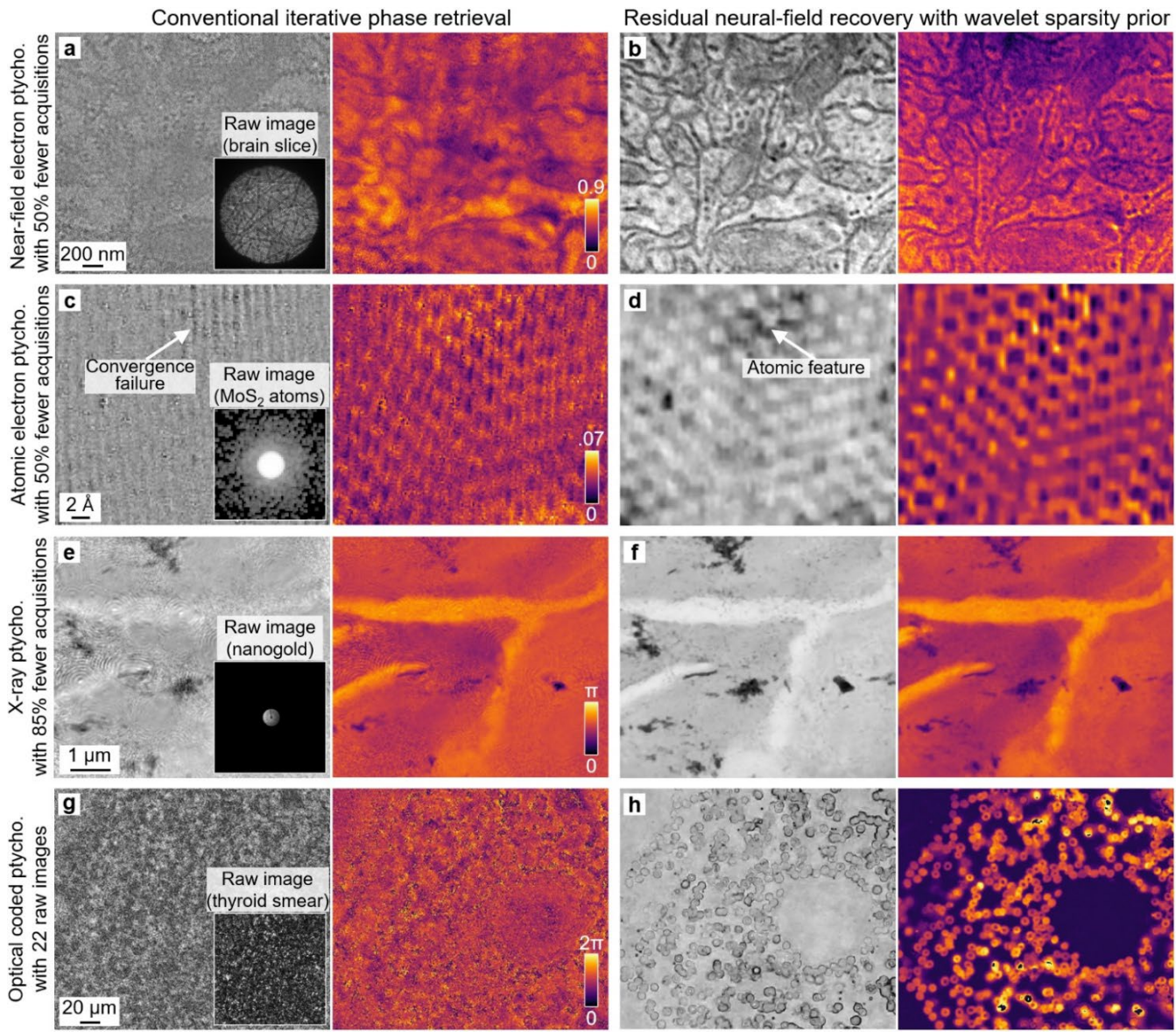

**Fig. 4 | Data-efficient reconstruction with wavelet sparsity constraints.** Comparison between conventional iterative phase retrieval (left column) and residual neural-field recovery with wavelet sparsity prior (right column) under severely reduced data acquisition. **a-b**, Near-field electron ptychography of brain tissue sections with 50% fewer acquisitions, showing amplitude and phase channels. **c-d**, Atomic-resolution electron ptychography of $MoS_2$ with 50% fewer acquisitions. **e-f**, Soft X-ray ptychography of nanogold sample with 85% fewer acquisitions. **g-h**, Optical coded ptychography of thyroid smear with only 22 raw images. Insets show representative raw diffraction patterns for each modality. Supplementary Figs. S6-S9 provide systematic comparisons across varying measurement redundancy levels for each modality.

For optical coded ptychography, we acquired lensless diffraction data from a thyroid smear specimen with pre-calibrated coded layer profiles (Methods). Using only 22 raw images (Fig. 4g-h), far below the typical hundreds of measurements, conventional approaches produced diffuse, noisy outputs with no discernible cellular morphology. The sparsity-constrained residual framework successfully reconstructed the thyroid smear specimen with subcellular detail and quantitative phase contrast, revealing individual cell boundaries and internal structures. In Supplementary Figs. S6-S9, we further conduct a systematic comparison against conventional iterative methods across varying measurement redundancy levels, confirming the superior data efficiency of our framework. Across all configurations, the combination of residual learning and wavelet sparsity enables robust reconstruction from measurements that would cause conventional iterative solvers to fail, significantly reducing acquisition time and radiation dose.

**Self correction via differentiable layers within the neural network**

A defining capability of our end-to-end differentiable framework is the joint optimization of experimental parameters alongside wavefield reconstruction. We demonstrated this self-correction capability by treating scan positions as learnable variables within the differentiable imaging model (Fig. 5). For near-field electron ptychography of brain tissue sections (Fig. 5a-c), reconstruction without position correction (Fig. 5a) exhibited blurring and reduced contrast in both amplitude and phase channels due to uncorrected positioning errors. Enabling position correction within the differentiable framework (Fig. 5b) improved reconstruction quality, revealing fine neural structures with enhanced sharpness. The scan position plot (Fig. 5c) shows the corrected positions (blue triangles) deviating systematically from the nominal positions (red circles), indicating that the framework identified and compensated for stage drift or calibration errors during acquisition.

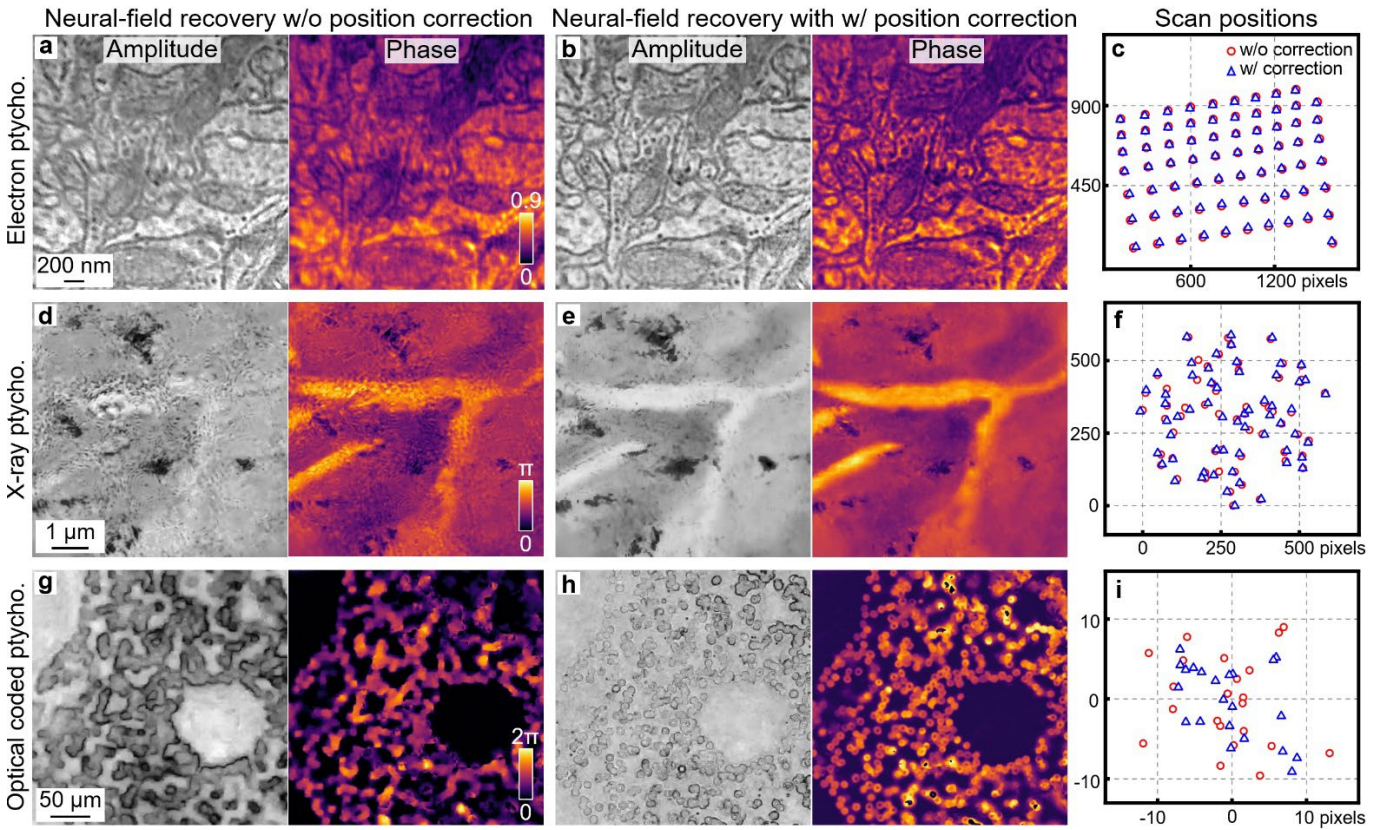

**Fig. 5 | Self-correction of scan positions via end-to-end differentiable physics.** Demonstration of automatic position correction across electron, X-ray, and optical ptychography. Left column: neural-field recovery without position correction. Middle column: neural-field recovery with position correction enabled within the differentiable imaging model. Right column: comparison of nominal scan positions (red circles) and corrected positions (blue triangles). **a-c**, Near-field electron ptychography of brain tissue sections, showing improved contrast and sharpness after position correction. **d-f**, Soft X-ray ptychography, demonstrating compensation for mechanical positioning uncertainties at synchrotron sources. **g-i**, Optical coded ptychography, revealing correction of random positioning jitter during coded sensor translation. The systematic deviations between nominal and corrected positions confirm that the framework successfully identifies and compensates for experimental positioning errors directly from diffraction data. Supplementary Fig. S10 demonstrates automatic calibration of additional experimental parameters including illumination wavelength and propagation distance. Supplementary Movie 1 visualizes the simultaneous refinement of propagation distance and scan positions during reconstruction.

For soft X-ray ptychography (Fig. 5d-f), we evaluated the framework's self-correction under challenging positioning conditions. Without position correction (Fig. 5d), the reconstruction exhibited artifacts and reduced phase contrast due to uncorrected positional deviations. With position correction enabled (Fig. 5e), the framework recovered cleaner amplitude and phase maps with improved structural definition. The position correction plot (Fig. 5f) reveals substantial deviations between nominal and corrected scan positions, demonstrating that the differentiable framework can successfully recover accurate positions even when initial estimates contain significant errors.

For optical coded ptychography (Fig. 5g-i), we further tested the robustness of position self-correction by introducing large perturbations to the stage feedback. Reconstruction without correction (Fig. 5g) produced noisy phase maps with reduced contrast, as expected when position errors remain uncompensated. With position correction enabled (Fig. 5h), the framework achieved clean phase reconstruction with well-defined cell boundaries and quantitative phase values despite the corrupted initial positions. The scatter plot (Fig. 5i) shows corrected positions compared to the nominal positions. Beyond scan position correction, the differentiable framework enables automatic calibration of other experimental parameters including illumination wavelength and propagation distance, as detailed in Supplementary Note 5, Supplementary Fig. S10, and Supplementary Movie 1.

These results demonstrate that embedding the imaging model as a differentiable layer enables automatic correction of system uncertainties directly from diffraction data, eliminating the need for interferometric position monitoring or post-acquisition position refinement algorithms. The same differentiable framework can be extended to jointly optimize other experimental parameters including aberration coefficients, as demonstrated in the Fourier ptychography experiments with learnable Zernike coefficients.

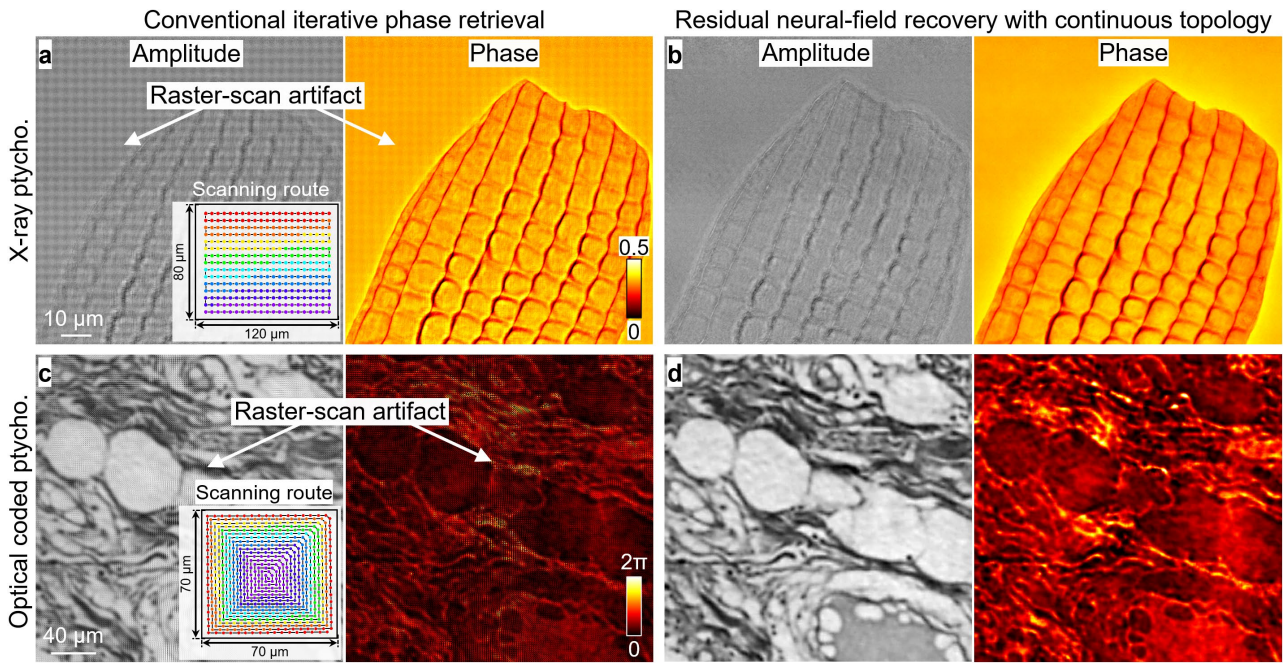

**Fig. 6| Suppression of raster-scan pathology via continuous neural topology.** Comparison of reconstructions using conventional iterative phase retrieval (left column) versus the proposed residual neural-field framework (right column) on data acquired with standard raster-scan trajectories. **a**, Conventional reconstruction of hard X-ray ptychography data (butterfly wing) using a raster scan pattern (inset). The output is corrupted by characteristic 'raster-grid pathology,' visible as periodic grid-like artifacts. **b**, Residual neural-field reconstruction of the same data. The continuous network topology inherently suppresses these periodic modes, yielding a clean, artifact-free image. **c**, Conventional reconstruction of optical coded ptychography data (biological tissue smear) using a raster scan, exhibiting periodic artifacts that degrades both amplitude and phase contrast. **d**, Corresponding neural-field recovery, which eliminates the artifacts to reveal clear cellular morphology and quantitative phase.

**Suppression of raster-grid pathology via continuous neural topology**

Beyond wavelength universality, the continuous topology of the neural field offers a fundamental advantage in mitigating scan-induced artifacts. Conventional pixel-based phaser retrieval solvers are frequently plagued by 'raster-grid pathology', a periodic reconstruction error characterized by grid artifacts[46]. This issue arises when the scanning periodicity aligns with the pixel grid, creating a periodic error potential that traps the optimizer in local minima. Consequently, experimentalists are often forced to adopt complex non-periodic scanning trajectories, such as Fermat spirals, which complicate stage control and limit scanning speed. In contrast, our framework inherently suppresses these artifacts. Because the object is parameterized as a continuous function rather than a discrete array, the network topology does not support the periodic modes associated with grid noise. As demonstrated in Fig. 6, our framework produces artifact-free, high-fidelity reconstructions from simple raster scans for both X-ray and optical modalities, whereas conventional algorithms applied to the same data succumb to severe grid artifacts (also refer to Supplementary Note 6, Supplementary Fig. S11, and Supplementary Movie 2). By breaking the dependency on randomized scanning, the framework re-enables the use of standard raster trajectories, significantly simplifying hardware control for high-throughput implementations.

## Discussion

We have introduced self-correcting residual neural fields as a unified framework for ptychography across electron, X-ray, and optical wavelengths. The framework addresses three fundamental challenges that have limited the practical deployment of neural-field-based phase retrieval: the need for complex-valued representations that respect wave physics, the difficulty of convergence without proper initialization priors, and the joint optimization of experimental parameters alongside neural-field reconstruction.

The adoption of holomorphic phasor activation, extending sinusoidal representation networks (SIREN) to the complex domain, proves essential for coherent diffraction imaging. Conventional neural fields that split complex wavefields into independent amplitude-phase or real-imaginary channels fundamentally cannot capture the intrinsic coupling between these quantities during wave propagation. Our complex-valued architecture preserves this coupling throughout the network, enabling more physically consistent representations. The phasor activation further provides a natural basis for representing oscillatory wavefields, analogous to how Fourier components describe wave phenomena. This design choice manifests in practical advantages: the network can represent high-frequency diffraction features without the spectral bias that limits conventional architectures, as demonstrated by the 244-nm resolution achieved in visible-light coded ptychography.

The residual learning formulation represents a departure from the *ab initio* reconstruction paradigm that dominates existing neural-field approaches. By anchoring optimization to physical priors derived from calibrated probes, known optical parameters, or preliminary reconstructions, we transform an ill-posed inverse problem into a well-conditioned one. The network learns only residual corrections rather than complete wavefields, focusing optimization capacity on the features that encode resolution rather than global structures that can be estimated from experimental parameters. This formulation also resolves the fundamental ambiguities inherent in joint object-probe retrieval: without proper initialization priors, phase accumulation in the object can be equivalently explained by probe shifts, or slowly varying phase can be absorbed into probe or coded surface profiles. The consistent failure of non-residual reconstruction across all tested configurations confirms that prior-constrained residual learning is essential for breaking these symmetry degeneracies.

Embedding the physical imaging model as a differentiable layer within the computational graph enables capabilities beyond conventional phase retrieval. By implementing Fourier transforms, Fresnel propagation, and probe-object interactions as differentiable operations, gradients flow from the loss function through the imaging physics to all learnable parameters. This architecture allows experimental variables that are conventionally treated as fixed inputs, including scan positions, diffraction distances, and aberration coefficients, to be jointly optimized during reconstruction. The self-correction of scan positions demonstrated across electron, X-ray, and optical modalities eliminates the need for interferometric position monitoring or separate position refinement algorithms. Similarly, the joint optimization of Zernike coefficients in Fourier ptychography enables automatic aberration correction without dedicated calibration procedures. This differentiable physics paradigm generalizes beyond the specific parameters demonstrated here: any quantity that influences the forward model and has a well-defined gradient can be incorporated as a learnable variable.

The incorporation of wavelet-domain sparsity constraints addresses the redundancy requirements that limit ptychographic throughput. Conventional phase retrieval algorithms typically require 50% or greater overlap between adjacent scan positions to ensure robust convergence, constraining acquisition speed and increasing radiation dose.

By exploiting the natural sparsity of residual signals in the wavelet domain, our framework achieves high-fidelity reconstruction from severely undersampled datasets: 50% reduction for electron ptychography, 85% for X-ray ptychography, and reconstruction from only 22 raw images for optical coded ptychography. This data efficiency is particularly valuable for dose-sensitive specimens, as demonstrated by the near-field electron ptychography of brain tissue sections revealing synaptic structures with half the conventional electron exposure.

The wavelength-agnostic nature of our framework distinguishes it from approaches tailored to specific imaging regimes. The same network architecture and hyperparameters successfully reconstruct atomic lattices in electron microscopy, nanoscale features in X-ray imaging, and subcellular structures in optical microscopy, spanning nearly six orders of magnitude in wavelength. This universality arises from the physics-informed design: the phasor activation encodes oscillatory behavior common to all coherent wavefields, the residual formulation leverages priors available across all ptychographic configurations, and the differentiable forward models accommodate the distinct propagation physics of each regime. The framework further generalizes across ptychographic modalities, successfully reconstructing conventional scanning, near-field, coded, and Fourier ptychography without modification to the core architecture.

Several avenues exist to extend this framework. Integration with multislice propagation models[62, 63] would enable reconstruction of optically thick specimens where the multiplicative approximation breaks down. Extension to ptychographic tomography[13, 64] would provide three-dimensional imaging capability by incorporating multiple illumination angles into the differentiable forward model. The continuous neural-field representation naturally accommodates non-Cartesian sampling geometries and could enable reconstruction from arbitrary scan patterns optimized for specific imaging objectives[65]. Finally, the differentiable physics paradigm could be extended to other computational imaging modalities beyond ptychography, including holography[66], synthetic aperture imaging[67], coherent diffraction imaging[68], and phase-contrast tomography[35], where similar challenges of phase retrieval and parameter calibration arise.

By unifying complex-valued neural representations, prior-constrained residual learning, and end-to-end differentiable physics, this framework overcomes limitations of both conventional iterative algorithms and existing neural-field approaches. The demonstrated capabilities, including record optical resolution, substantial data reduction, and automatic parameter calibration, establish a path toward high-throughput, dose-efficient nanoscopy across the electromagnetic spectrum.

## Methods
### Residual neural-field architecture and implementation
We implemented the residual neural field framework in PyTorch with Compute Unified Device Architecture (CUDA) acceleration. For multiresolution hash encoding, the spatial coordinate $r$ is mapped to a feature vector $f(r)$ by concatenating interpolated features from $L$ resolution levels:

$$f(r) = \bigoplus_{l=1}^{L} \mathrm{interp}(\mathcal{T}_l[h_l(r)]) \tag{5}$$

where $\mathcal{T}_l$ denotes the learnable feature table at resolution level $l$, $h_l$ is the spatial hashing function, $\mathrm{interp}(\cdot)$ represents bilinear interpolation, and $\oplus$ denotes vector concatenation. We use $L = 16$ resolution levels with $F = 2$ features per level, hash table size $2^{19}$, base resolution 16, and per-level scale factor 1.5, yielding a 32-dimensional feature vector for each spatial coordinate. The complex-valued MLP consists of one hidden layer with 64 complex neurons (equivalent to 128 real-valued neurons in representational capacity). The frequency parameter $\omega$ in the phasor activation (Eq. 1) is typically ranging from 1 to 20 for the input layer and fixed at 1 for the hidden layer across all configurations. Separate neural field networks $\mathcal{F}_{\theta_O}$ and $\mathcal{F}_{\theta_P}$ are instantiated to represent the complex object and probe residuals as defined in Eqs. (2) and (3). The differentiable forward model implements the imaging process for ptychographic modalities across electron to optical wavelengths, including conventional, near-field, coded, and Fourier configurations. The network is trained by minimizing a composite loss combining data fidelity with wavelet-domain sparsity regularization:

$$\mathcal{L}oss = \frac{1}{B} \sum_{i \in \mathrm{batch}} \mathcal{L}_{\mathrm{data}}(J_i, I_i) + \mu \cdot \frac{\mathcal{L}_{\mathrm{wavelet}}(O)}{N}, \tag{6}$$

where $J_i$ is the measured intensity, $I_i$ is the estimated intensity, $B$ is the batch size, and $N$ is the number of pixels in the object. For the data fidelity term, we use Huber loss (smooth L1) for electron ptychography and gradient-domain loss for optical and X-ray modalities, as the latter provides robustness against low-frequency background variations common in these regimes. The wavelet sparsity term $\mathcal{L}_{\text{wavelet}}$ imposes $\ell_1$ constraints on the Haar wavelet subbands of the complex object $O$. The Haar wavelet decomposition is computed efficiently through down-sampling operations on the four quadrants of the image. The regularization weight $\mu$ was set to 0.1 for the experiments presented.

Training was performed using the AdamW optimizer[54] with initial learning rate typically set to $10^{-3}$, momentum parameters $\beta_1 = 0.9$ and $\beta_2 = 0.999$, and weight decay $10^{-5}$. Each epoch randomly permutes the frame indices. Training typically converges within 10 epochs. For joint parameter optimization experiments (Fig. 5), scan positions $(x_i, y_i)$ are assigned separate learning rates (typically 0.1) to account for their different scales relative to neural network weights. For baseline comparisons in Supplementary Figs. 6-9, conventional iterative phase retrieval was performed using the extended ptychographical iterative engine (ePIE)[15], weighted average of sequential projections (WASP)[17], quasi-Newton[18], and least-squares solver[19]. These conventional methods used identical input data, initialization priors, and scan position information to ensure fair comparison.

**Brain tissue sample for electron ptychography**
We prepared the brain tissue sections following the established protocols for serial section electron microscopy[69]. Adult mice were anesthetized with sodium pentobarbital and transcardially perfused with 2% paraformaldehyde and 2.5% glutaraldehyde in 0.1 M cacodylate buffer (pH 7.4). Brains were extracted and post-fixed overnight at 4°C in the same fixative solution. Tissue blocks containing the region of interest were dissected and processed for electron microscopy as previously described[69]. Briefly, samples were rinsed in 0.1 M cacodylate buffer, post-fixed in 1% osmium tetroxide with 1.5% potassium ferrocyanide for 1 hour followed by 1% osmium tetroxide for 1 hour, dehydrated through an ascending series of ethanol dilutions (50%, 70%, 90%, and 100%) containing 1.5% uranyl acetate, and flat embedded in LX-112 epoxy resin (Ladd Research Industries). Blocks were trimmed to expose the target region, and ultrathin sections (45-70 nm) were cut using a Leica UC7 ultramicrotome and collected on pioloform-coated slot grids. This preparation protocol produces excellent ultrastructural preservation with intact membranes and strong contrast, enabling visualization of synaptic connectivity, dendritic spines, axon terminals, and subcellular organelles.

**Near-field electron ptychography for brain imaging**
We performed the near-field electron ptychography experiments on a Titan HOLO electron microscope (Thermo Fisher Scientific) operated at 80 keV. An amplitude diffuser composed of 350-nm-wide randomly arranged tracks within a 50-μm silicon nitride aperture was positioned in the condenser aperture plane. This configuration projects a structured illumination pattern onto the sample while making efficient use of electron dose, as there is no masking by post-specimen apertures. The microscope was equipped with a Gatan K2 electron detector (3840 × 3712 pixels, 5 μm pitch). For data acquisition, the sample was translated using stepper motor stages programmed via Digital Micrograph scripts. The condenser lenses were tuned to project the structured light onto the sample with a diameter of approximately 2 μm. The projection lenses were adjusted to provide appropriate defocus (approximately 390 μm) for near-field diffraction pattern formation.

**Lensless coded ptychography**
We performed lensless coded ptychography experiments using a 405 nm laser diode (LP405C1, Thorlabs). The optical setup consisted of a coded sensor configuration where a coded surface was fabricated on the coverglass of the image sensor (AR2020, ON Semiconductor). The coded surface transmission profile was pre-calibrated using a weakly scattering reference specimen (blood smear) prior to sample measurements[10]. For data acquisition, the coded sensor was translated through a grid of positions while recording near-field diffraction patterns at each location. The USAF resolution target (Fig. 3) and $Na_2CO_3$ crystal samples were imaged with this configuration, demonstrating resolution down to 244-nm linewidths with visible light, the best visible-light resolution achieved without lenses.

**Lens-based Fourier ptychography**

We conducted Fourier ptychography experiments using an inverted microscope with a programmable LED array for angle-varied illumination at 529 nm wavelength. Unlike conventional Fourier ptychography, we used a spatially-coded detection scheme by placing a coded surface in the image plane before the detector. This configuration provides a uniform phase transfer function enabling true quantitative phase imaging, overcoming the inability of regular Fourier ptychography to recover slowly varying phase information. The coded surface and pupil function priors were calibrated using a blood smear as a reference specimen. For reconstruction, the microscope pupil was parameterized using Zernike polynomial coefficients within the differentiable imaging model:

$$Pupil(\boldsymbol{k}) = A(\boldsymbol{k}) \cdot \exp\left(i \cdot \sum_{n,m} c_{n,m} Z_n^m(\boldsymbol{k})\right), \quad (7)$$

where $\boldsymbol{k} = (k_x, k_y)$ denotes the spatial frequency coordinates in the pupil plane, $A(\boldsymbol{k})$ represents the binary aperture function determined by the numerical aperture, $Z_n^m(\boldsymbol{k})$ are the Zernike polynomials with radial order $n$ and azimuthal frequency $m$, and $c_{n,m}$ are the learnable aberration coefficients. We included 10 Zernike terms corresponding to defocus, astigmatism, coma, trefoil, and spherical aberration. This parameterization enables joint optimization of aberration correction alongside wavefield reconstruction. Cystine crystal samples were imaged with this configuration.

**Atomic-resolution electron ptychography**

To validate the framework at atomic resolution, we utilized a dataset of a molybdenum disulfide ($MoS_2$) monolayer[3]. The data was acquired using an aberration-corrected FEI Titan Themis microscope operated at 80 kV (λ = 4.2 pm). The dataset consists of a 4D-STEM scan with a convergence semi-angle of 21.4 mrad. For our reconstruction, we utilized a central crop of the diffraction patterns to match the field of view of interest. The probe prior was initialized using the theoretical aperture function calculated from the convergence angle and defocus estimate.

**Hard and soft X-ray ptychography**

For hard X-ray ptychography, we utilized a benchmark dataset of a butterfly scale[61]. The measurements were acquired using a photon energy of 9.7 keV (λ = 1.3 Å). For soft X-ray ptychography, we utilized the nanogold dataset collected in the soft X-ray regime (0.71 keV, λ = 1.75 nm)[61]. For both datasets, we reduce the data by subsampling the scan positions to evaluate the robustness of the residual neural field framework under sparse sampling conditions.

**Open-source implementation and Jupyter notebooks**

We have provided the complete framework and experimental datasets as an open-source package. The implementation, developed in PyTorch with CUDA acceleration, encompasses the unified residual neural-field architecture and end-to-end differentiable forward models for electron, X-ray, and optical ptychography. We provide ready-to-run Jupyter notebooks (Recovery.ipynb) corresponding to the four key demonstrations in this work. A detailed description of the codebase structure, software dependencies, hardware requirements, and direct links to the Zenodo repository containing both the source code and datasets are provided in Supplementary Note 7.

**Data availability**

The experimental datasets that accompany the code implementations are available in Supplementary Note 7 with a Zenodo repository link.

**Code availability**

The code implementations for space-time neural field reconstruction and associated analysis scripts are available in Supplementary Note 7 with a Zenodo repository link.

**Acknowledgments**

This work was partially supported by the Department of Energy SC0025582 and the National Institute of Health R01-EB034744. The content of the article does not necessarily reflect the position or policy of the US government, and no official endorsement should be inferred. Q. Z. acknowledges the support of G. E. fellowship. We thank Dr. Linnaea

Ostroff from the Department of Physiology and Neurobiology at UConn for preparing the thin brain slides for near-field electron ptychography.

**Author contributions**
G. Z. conceived the original concept of residual neural-field ptychography and supervised the project. Q. Z. and Z. H. developed the neural field framework for electron, X-ray, and optical ptychography. P. H. L, R. E. D., and A. M. performed the near-field electron ptychography experiments. Q. Z., Z. H., and R. W. performed the optical experiments with lensless coded ptychography and lens-based Fourier ptychography. Q. Z., Z. H., R. W., T. W., and Q. M. prepared the display items and Supplementary Movies. Q. Z. prepared the Supplementary Information. All authors participated in the discussion and interpretation of the results. Q. Z. and G. Z. wrote the manuscript with the input from all authors.

**Competing interests**
G. Z. is a named inventor of a related patent application. Other authors declare no competing interests.

**Supplementary Information**
**Supplementary Figures S1-S12:**
Supplementary Fig. S1 | Experimental configurations and forward models for different ptychographic modalities.
Supplementary Fig. S2 | Comparative performance of complex-valued activations in ptychographic neural field reconstruction.
Supplementary Fig. S3 | Robustness of phasor activation performance across different frequency parameters.
Supplementary Fig. S4 | Optimizer-dependent reconstruction quality and convergence characteristics across different ptychographic modalities.
Supplementary Fig. S5 | Effect of wavelet sparsity regularization across ptychographic modalities.
Supplementary Fig. S6 | Near-field electron ptychography reconstruction comparison with varying measurement redundancy.
Supplementary Fig. S7 | Electron ptychography reconstruction comparison with varying measurement redundancy.
Supplementary Fig. S8 | X-ray ptychography reconstruction with varying measurement redundancy.
Supplementary Fig. S9 | Coded ptychography reconstruction with varying measurement redundancy.
Supplementary Fig. S10 | Experimental parameter calibration via differentiable reconstruction.
Supplementary Fig. S11 | Raster scan artifacts in conventional phase retrieval algorithms.

**Supplementary Notes 1-7:**
Supplementary Note 1 | Ptychographic modalities and forward models.
Supplementary Note 2 | Comparison of complex activations in ptychographic neural fields.
Supplementary Note 3 | Optimizer choice for residual neural-field ptychography.
Supplementary Note 4 | Wavelet sparsity regularization for neural field reconstruction.
Supplementary Note 5 | Automatic parameter calibration via differentiable reconstruction.
Supplementary Note 6 | Suppression of raster-grid pathology.
Supplementary Note 7 | Open-source dataset description.

**Supplementary Movies 1-2:**
Supplementary Movie 1 | Automatic parameter calibration in coded ptychography via differentiable reconstruction.
Supplementary Movie 2 | Suppression of raster-grid pathology via continuous neural representation.